\documentclass[onecolumn,draftclsnofoot]{IEEEtran}
\usepackage{amsmath}
\usepackage{graphicx}
\usepackage{caption2}
\usepackage{amsthm}
\begin{document}
\title{DoF Analysis of the K-user MISO Broadcast Channel with Alternating CSIT}
\author{{Borzoo Rassouli, Chenxi Hao and Bruno Clerckx} %
\thanks{Borzoo Rassouli, Chenxi Hao and Bruno Clerckx are with the Communication and Signal Processing group of Department of Electrical and Electronics,
Imperial College London, email: \{b.rassouli12; chenxi.hao10; b.clerckx\}@imperial.ac.uk}
\thanks{This work was partially supported by the Seventh Framework Programme for Research of the European Commission under grant number HARP-318489.}}
\maketitle
\begin{abstract}
%\boldmath
We consider a $K$-user multiple-input single-output (MISO) broadcast channel (BC) where the channel state information (CSI) of user $i(i=1,2,\ldots,K)$ may be either perfect (P), delayed (D) or not known (N) at the transmitter with probabilities $\lambda_P^i$, $\lambda_D^i$ and $\lambda_N^i$, respectively. In this channel, according to the three possible CSIT for each user, joint CSIT of the $K$ users could have at most $3^K$ realizations. Although the results by Tandon et al. show that the Degrees of Freedom (DoF) region for the two user MISO BC with symmetric marginal probabilities (i.e., $\lambda_Q^i=\lambda_Q \forall i\in \{1,2,\ldots,K\}, Q\in \{P,D,N\}$) depends only on the marginal probabilities, we show that this interesting result does not hold in general when the number of users is more than two. In other words, the DoF region is a function of the \textit{CSIT pattern}, or equivalently, all the joint probabilities. In this paper, given the marginal probabilities of CSIT, we derive an outer bound for the DoF region of the $K$-user MISO BC. Subsequently, the achievability of these outer bounds are considered in certain scenarios. Finally, we show the dependence of the DoF region on the joint probabilities.
\end{abstract}


\section{Introduction}
In contrast to the point to point multiple-input multiple-output (MIMO) communication where the channel state information at the transmitter (CSIT) does not affect the multiplexing gain, in a multiple-input single-output (MISO) broadcast channel (BC), knowledge of CSIT is crucial for interference mitigation and beamforming purposes \cite{Bruno}. However, the assumption of perfect CSIT may not always be true in practice due to channel estimation and feedback latency. Therefore, the idea of communication under some sort of imperfection in CSIT has gained more attention recently. The so called MAT algorithm was presented in \cite{MAT} where it was shown that in terms of the degrees of freedom, even an outdated CSIT can result in significant performance improvement in comparison to the case with no CSIT. Assuming correlation between the feedback information and current channel state (e.g., when the feedback latency is smaller than the coherence time of the channel), the authors in \cite{Gesbert} and \cite{Gou12} consider the degrees of freedom in a time correlated MISO BC which is shown to be a combination of zero forcing beamforming (ZFBF) and MAT algorithm. Following these works, the general case of mixed CSIT and the $K$-user MISO BC with time correlated delayed CSIT are discussed in \cite{Chen12a} and \cite{xinping_Kuser}, respectively. While all these works consider the concept of delayed CSIT in time domain, \cite{Chenxi} and \cite{Hao} deal with the DoF region and its achievable schemes in a frequency correlated MISO BC where there is no delayed CSIT but imperfect CSIT across subbands, which is more inline with practical systems as Long Term Evolution (LTE) \cite{Bruno}. The most relevant article to this paper is the work done in \cite{Tandon} where the synergistic benefits of alternating CSIT over fixed CSIT was presented in a two user MISO BC with two transmit antennas. The converse in \cite{Tandon} is based on the idea of assigning artificial receivers to the users whose observations are (statistically) equivalent to the corresponding user when CSIT is (not) perfect. However, whether this brilliant approach could be generalized to the scenarios with more than two transmit antennas and two users is unknown. Therefore, for such scenarios, it becomes necessary to check other ways to find the fundamental limits of the system. To the best of our knowledge, this is the first paper in the literature addressing the general $K$-user MISO BC with alternating CSIT. To this end, our contributions are as follows.
\begin{itemize}
  \item Given the marginal probabilities of CSIT in a $K$-user MISO BC, we derive an outer bound for the DoF region where the proof is based on finding upper bounds for a certain difference between entropies and is inspired by \cite{Hao} and the results in \cite{extremal}.
  \item We investigate the achievability and tightness of the outer bounds. Several achievable schemes are introduced and shown to achieve the corner points of the DoF region in some scenarios, therefore proving that the outer bounds are optimal bounds in those scenarios.
  \item Finally, we provide an example which proves that in contrast to the results of \cite{Tandon} for the two user BC, the DoF region of the $K$-user MISO BC ($K\geq3$) is not only a function of marginal probabilities in general.
\end{itemize}

The paper is organized as follows. In section \ref{s2} the system model and preliminaries are presented. The main result of this paper is provided in section \ref{s3} as a theorem. The proof and tightness of the outerbounds will be discussed in section \ref{ss4} and \ref{s55} , respectively. Section \ref{sh} shows that the DoF region depends on the joint CSIT probabilities in general, and section \ref{s7} concludes the paper.

Throughout the paper, vectors are shown in bold lower case while matrices are written in upper case. $CN(\textbf{0},\mathbf{\Sigma})$ is the circularly symmetric complex Gaussian distribution with covariance matrix $\mathbf{\Sigma}$. $f\sim O(\log P)$ is equivalent to $\lim_{P\to \infty}\frac{f}{\log P}=0$. $X_i^n=\{X(i),X(i+1),\ldots,X(n)\}$ is the time extension of random variable $X$ and when $i=1$, it is dropped for simplicity (i.e., written as $X^n$). $(.)^T$ and $(.)^H$ denote the transpose and conjugate transpose, respectively. Both of the terms upper bound and outer bound, used in this paper, have almost similar meanings with a slight difference; while the former is only used for scalars, the latter is a more general term used for multidimensional regions and could be defined by (in)finite number of upper bounds. Finally, Let $S_1$ and $S_2$ be two sets of inequalities defining the regions $D_1$ and $D_2$, respectively, and assume the region $D$ is defined by the set of inequalities $S=S_1\cup S_2$ or equivalently $D=D_1\cap D_2$. The set of inequalities $S_1$ is called inactive (or redundant) in defining $D$ when $D_2\subset D_1$.
\section{System Model}\label{s2}
We consider a MISO BC, in which a base station with $M$ antennas sends independent messages $W_1,\ldots,W_K$ to $K$ single-antenna users ($M\geq K$). In a flat fading scenario, the discrete-time baseband received signal of user $k$ at time instant $n$ can be written as
\begin{equation}
  y_k(n)=\textbf{\textit{h}}_{k}^{H}(n)\textbf{\textit{x}}(n) + w_k(n) \ \ \ ,\ \ \ k=1,2,\ldots,K
\end{equation}
where $\textit{\textbf{x}}(n)\in C^{(M\times1)}$ is the transmitted signal at time instant $n$ satisfying the power constraint $E\left[\|\textit{\textbf{x}}\|^2\right]\leq P$, and $w_k(n)$ is the additive noise with distribution $CN(0,1)$. The channel vector of user $k$ has the distribution $CN(\textbf{0},\textbf{I})$ and is i.i.d. over time and users. Also, let $H(n)={[\textbf{\textit{h}}_1(n),\ldots,\textbf{\textit{h}}_K(n)]}^H$ and $H^n=\{H(1),\ldots,H(n)\}$. We assume global perfect Channel State Information at Receiver (CSIR)  i.e., at time instant $n$, all users have perfect knowledge of $H^n$.

The rate tuple $(R_1,R_2,\ldots,R_K)$ is achievable if the probability of error in decoding $W_i$ at user $i (i=1,\ldots,K)$ can be made arbitrarily small with sufficiently large coding length. Analysis of the capacity region, which is the set of all achievable rate tuples, is not always tractable. Instead, we consider the DoF region, which is a simpler metric independent of the transmit power, and is defined as $\{(d_1,\ldots,d_K)|d_i=\lim_{P\to \infty}\frac{R_i}{\log P} \forall i,R_i\}$. At very high SNRs, the effect of noise can be neglected and what remains is the interference caused by other users' signals. Therefore, the DoF region could also be interpreted as the set constructed by the number of interference-free private data streams that each user receives per channel use.
\begin{figure}[t]
  \centering
  \includegraphics[width=10cm]{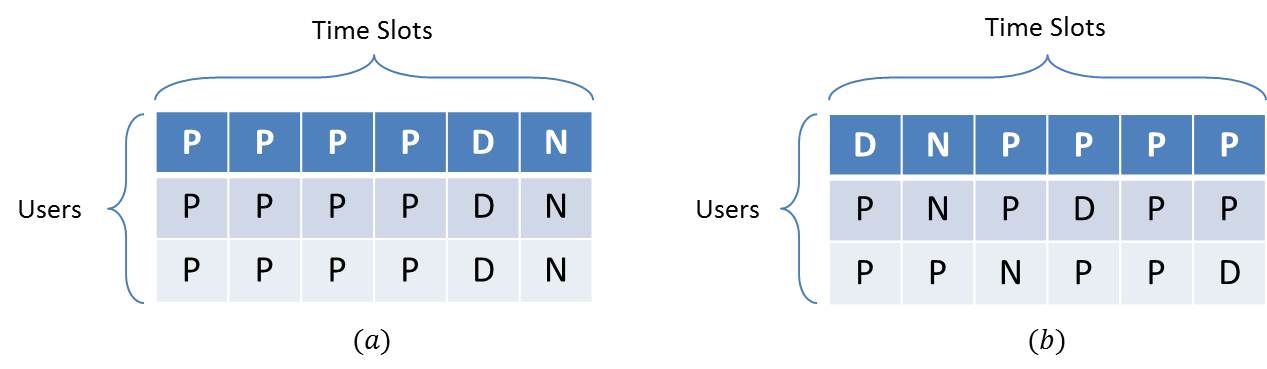}\\
  \caption{Two different CSIT patterns both having the same marginal probabilities ($\lambda_P=4\lambda_D=4\lambda_N=\frac{2}{3}$)}\label{fig7}
\end{figure}

The CSIT model used in this paper is the same as that in \cite{Tandon} i.e., at some time instants the transmitter has a Perfect (P) instantaneous knowledge of the CSI of a particular user, whereas at some time instants it receives the CSI with Delay (D) and finally, at some time instants the CSI of the user is Not known (N) at the transmitter. When there is delayed CSIT, we assume that the feedback delay is larger than the coherence time of the channel making the feedback information completely independent of the current channel state. In this configuration, the joint CSIT of all the $K$ users has at most $3^K$ states. For example, in a 3 user MISO BC, they will be $PPP,PPD,PPN,PDP,\ldots$  with corresponding probabilities $\lambda_{PPP}, \lambda_{PPD},\lambda_{PPN},\lambda_{PDP},\ldots$\ . A scenario is symmetric when the marginal CSIT probabilities are the same across the users (i.e., $\lambda_Q^i=\lambda_Q \forall i\in \{1,2,\ldots,K\}, Q\in \{P,D,N\}$) and asymmetric otherwise. For the symmetric case, it was shown in \cite{Tandon} that the DoF region for the $2$ user BC with $2$ transmit antennas at the base station is only a function of marginal probabilities. We show that this interesting result also holds for an arbitrary number of antennas ($>2$) in the two user MISO BC, however it does not hold for the general $K(>2)$ user BC and the DoF region is a function of the CSIT pattern (i.e., a function of all $3^K$ joint state probabilities.) By CSIT pattern we refer to the knowledge of CSIT represented in a space-time matrix where the rows and columns represent users and time slots, respectively. Figure \ref{fig7} shows two different symmetric CSIT patterns where both have the same marginal probabilities. For example, in pattern $(b)$, the transmitter knows the channels of users 2 and 3 perfectly at time slot 1 and has no information about the channel of user 1. The CSI of user 1 will be known in the next time slot (i.e., time slot 2) due to feedback delay and is completely independent of the channel in time slot 2.
\begin{figure}[t]
  \centering
  \includegraphics[width=10cm]{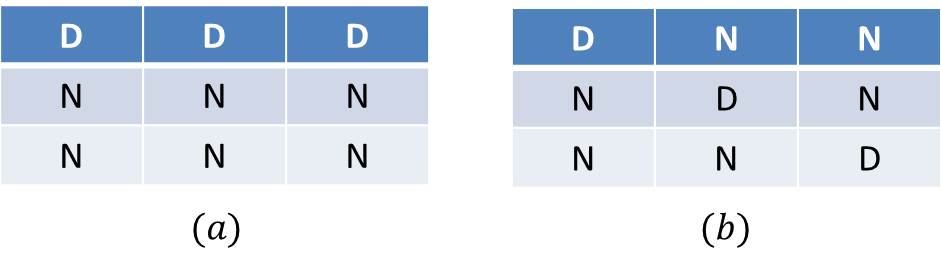}\\
  \caption{An example of the fixed and alternating CSIT in a 3-uer MISO BC $(a)$ fixed CSIT $(b)$ alternating CSIT}\label{f6}
\end{figure}

\begin{figure}
  \centering
  \includegraphics[width=15cm]{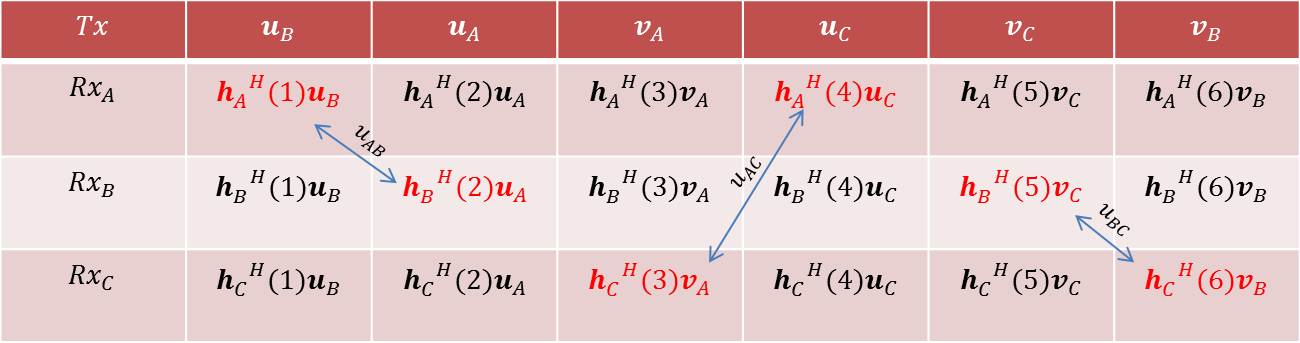}\\
  \caption{Transmission scheme for the example of alternating CSIT pattern}\label{f7}
\end{figure}

The synergistic benefit of alternating CSIT over fixed CSIT was shown in \cite{Tandon} for the 2-user MISO BC. It is interesting to check whether it holds for the general $K$-user MISO BC. For example, consider a 3-user MISO BC in which the transmitter has always delayed CSI of one user and no CSI of the remaining users like the pattern shown in figure \ref{f6} $(a)$. Now, what happens if the CSIT alternates among the users as in pattern $(b)$? Is it beneficial? The answer is yes. According to the results of this paper (see the theorem in section \ref{s3}), the sum DoF of the fixed CSIT (i.e., pattern $(a)$) has an upper bound of 1, and since it is simply achievable, the upper bound is tight. However, a sum DoF of $\frac{24}{17}(>1)$ is achievable for the alternating case (pattern $(b)$) as follows. Let $\textit{\textbf{u}}_X$ and $\textit{\textbf{v}}_X$ be two complex vectors in $C^{2\times 1}$ where each of them contains two symbols from two independently encoded Gaussian codewords intended for user $X(= A,B$ or $C).$ For brevity, the transmission scheme is shown in figure \ref{f7} where in the first row the transmitted vectors in consecutive time slots are shown, and the remaining three rows show the received signal at the corresponding receiver. In conjunction with pattern $(b)$ in figure \ref{f6}, the received signal of the users that feed back their channel is shown in red. After the end of time slot 2, the transmitter knows both $\textit{\textbf{h}}_A^H(1)$ and $\textit{\textbf{h}}_B^H(2)$, and if it sends the scalar symbol $u_{AB}=\textit{\textbf{h}}_A^H(1)\textit{\textbf{u}}_B+\textit{\textbf{h}}_B^H(2)\textit{\textbf{u}}_A$ to both receivers $A$ and $B$, both of them can decode their intended vectors, by means of interference cancellation. Such a message ($u_{AB}$) is called an order-2 message, since it is intended for two receivers \cite{MAT}. Therefore, for successful detection of 12 symbols, 3 order-2 messages ($u_{AB}, u_{AC}$ and $u_{BC}$) must be sent. According to \cite{MAT}, in a 3-user scenario, 6 order-2 messages can be sent in a frame of 5 time slots where the first three time slots of the frame look the same as pattern $(b)$ in figure \ref{f6} and in the next two time slots, the transmitter requires no CSIT. Hence, 12 symbols will be successfully decoded in $6+\frac{3}{\frac{6}{5}}$ time slots which results in an achievable sum DoF of $\frac{12}{6+\frac{3}{\frac{6}{5}}}(=\frac{24}{17})$ which is obviously greater than the sum DoF of the fixed CSIT scenario. Therefore, CSIT alternation is also beneficial in the $K$-user scenario.

The main result of this paper is that given the marginal probabilities of CSIT, an outer bound for the DoF region is provided regardless of the CSIT pattern and its achievability is considered in some scenarios with specific patterns. For example, according to the results of this paper, given the marginal probabilities as in figure \ref{fig7} ($\lambda_P=\frac{2}{3}, \lambda_D=\lambda_N=\frac{1}{6}$), the sum DoF has the upper bound of $\frac{28}{11}$. Actually, it is optimal for the CSIT pattern in $(a)$, since it is achievable, however whether it is also tight for the pattern in $(b)$ is unknown. It is our conjecture that this is not a tight bound for the latter, though having the same marginal probabilities as in $(a)$. This dependency on CSIT pattern, which will be discussed in section \ref{sh}, is equivalent to having the optimal DoF region as a function of $\lambda_{DDP},\lambda_{PNN},\ldots$ in such a way that they do not add up to produce only the marginal probabilities, in contrast to the results of \cite{Tandon} for the 2-user case.

\section{Main results}\label{s3}

\textit{Theorem}. Let $O^K$ be the outer bound for the DoF region of the $K$ user MISO BC with $M$ transmit antennas at the transmitter ($M\geq K$). Given the marginal probabilities of CSIT for user $i$ (which can be any two of $\lambda_P^i, \lambda_D^i$ and $\lambda_N^i$, since $\lambda_P^i + \lambda_D^i + \lambda_N^i = 1$),
\begin{equation} \label{theorem}
  O^K=\left\{(d_1,d_2,...,d_K)|\sum_{i=1}^j\frac{d_{\pi^j(i)}}{i}\leq 1 + \sum_{i=2}^j\frac{\sum_{r=1}^{i-1}\lambda_P^{\pi^j(r)}}{i(i-1)}\ ,\  \sum_{i=1}^j d_{\pi^j(i)} \leq 1 + \sum_{i=1}^{j-1}(\lambda_P^{\psi_{\pi^j}(i)} + \lambda_D^{\psi_{\pi^j}(i)}), \forall \pi^j, j=1,2,...,K\right\}
\end{equation}
where $\pi^j(.)$ is an arbitrary permutation of size $j$ over the indices $(1,2,\ldots,K)$, and $\psi_{\pi^j}(.)$ is a permutation of $\pi^j$ satisfying
\begin{equation}\label{const}
  (\lambda_P^{\psi_{\pi^j}(i)} + \lambda_D^{\psi_{\pi^j}(i)})\leq (\lambda_P^{\psi_{\pi^j}(i+1)} + \lambda_D^{\psi_{\pi^j}(i+1)})\ \ \ \ \ ,\ \ \ \ i=1,2,\ldots,j-1.
\end{equation}
For the symmetric scenario, $O^K$ is simplified as
\begin{equation}
  O^K=\left\{(d_1,d_2,...,d_K)|\sum_{i=1}^j\frac{d_{\pi^j(i)}}{i}\leq 1 + \lambda_P\sum_{i=2}^j\frac{1}{i}\ ,\  \sum_{i=1}^j d_{\pi^j(i)} \leq 1 + (j - 1)(\lambda_P + \lambda_D), \forall \pi^j, j=1,2,...,K\right\}.
\end{equation}
It is important to note that these outer bounds hold regardless of the CSIT patterns and are only a function of marginal probabilities. The proof is provided in the next section.
\section{Proof of the theorem}\label{ss4}
The structure of the proof could be briefly itemized as follows.
\begin{itemize}
  \item Applying some sort of improvement to the channel.
  \item The usage of Fano's inequality.
  \item Application of the \textit{Csisz\'{a}r sum identity} \cite{network_info} as in \cite{Hao} to change the difference between vector entropies into the sum of the component-wise entropy differences.
  \item Finding an upper bound for these entropy differences by application of two provided lemmas.
\end{itemize}
Having an outer bound for the DoF region of the general $K$-user BC ($M\geq K$), it is obvious that each subset of users with cardinality $j$ ($j<K$) should satisfy the outer bound for the $j$-user BC ($O^j$). Therefore, we only consider the proof of the inequalities involving all the $K$ users. For simplicity, we show the inequalities for the identity permutation (i.e., $\pi^K(i)=i$) while the results could be easily extended to any other arbitrary permutation.
\subsection{Proof of $\sum_{i=1}^K\frac{d_i}{i}\leq 1 + \sum_{i=2}^K\frac{\sum_{r=1}^{i-1}\lambda_P^r}{i(i-1)}$}
First, we improve the channel by giving the message and observation of user $i$ to users $i+1$ to $K$\ ($i=1,\ldots,K-1$). Hence, from Fano's inequality,
\begin{equation}\label{Fano}
  nR_i \leq I(W_i;Y_1^n,\ldots,Y_i^n|W_1,\ldots,W_{i-1},H^n)+\epsilon_n
\end{equation}
where $W_0=\emptyset$. This improvement does not decrease the capacity region, meaning that the capacity region of the original channel is a subset of this improved channel. Also, by this improvement, channel input and outputs (i.e., the enhanced observations of users) form a Markov chain which results in a degraded broadcast channel \cite{Cover}. Therefore, according to \cite{Elgamal}, since feedback does not increase the capacity of degraded broadcast channels, we can ignore the delayed CSIT (D) and replace them with No CSIT (N). In other words, at time instant $n$, knowledge of the CSI up to time instant $n-1$ is not beneficial in a physically degraded BC. Therefore, it is equivalent to having the channel of user $i$ perfectly known with probability $\lambda_P^i$ and not known otherwise. It is important to note that although the channel has become physically degraded, the perfect CSIT (P) cannot be replaced with No CSIT (N), since (P) means that at time instant $n$ the current state of the channel is known to the transmitter perfectly which enables it to know the received signal within noise level (i.e., the results of \cite{Elgamal} cannot be applied in this case.) From now on, we ignore the term $\epsilon_n$ for simplicity and write
\begin{align}
\sum_{i=1}^K\frac{nR_i}{i}  &\leq  \sum_{i=1}^K\frac{I(W_i;Y_1^n,\ldots,Y_i^n|W_1,\ldots,W_{i-1},H^n)}{i}  \\
 &\leq h(Y_1^n|H^n) +\sum_{i=2}^K\left[\frac{h(Y_1^n,\ldots,Y_i^n|W_1,\ldots,W_{i-1},H^n)}{i}- \right.
   \left. \frac{h(Y_1^n,\ldots,Y_{i-1}^n|W_1,\ldots,W_{i-1},H^n)}{i-1}\right] \nonumber \\&\ \ \ + nO(\log P)\label{e1}
\end{align}
where $Y_0=\emptyset$ and we have used the fact that
\begin{equation}
  \frac{h(Y_1^n,\ldots,Y_K^n|W_1,\ldots,W_K,H^n)}{nK}\sim O(\log P). \nonumber
\end{equation}
since with the knowledge of $W_1,\ldots,W_K,H^n$, the observations $Y_1^n,\ldots,Y_K^n$ can be reconstructed within noise distortion.
From the chain rule of entropies, each of the terms in the summation in (\ref{e1}) can be written as
\begin{multline}\label{eq}
\frac{h(Y_1^n,\ldots,Y_i^n|W_1,\ldots,W_{i-1},H^n)}{i}
-\frac{h(Y_1^n,\ldots,Y_{i-1}^n|W_1,\ldots,W_{i-1},H^n)}{i-1} \\
=\sum_{j=1}^n\left[\frac{h(Y_1(j),\ldots,Y_i(j)|W_1,\ldots,W_{i-1},Y_1^{j-1},\ldots,Y_i^{j-1},H^j)}{i}\right.
\\ \left.-\frac{h(Y_1(j),\ldots,Y_{i-1}(j)|W_1,\ldots,W_{i-1},Y_1^{j-1},\ldots,Y_{i-1}^{j-1},H^j)}{i-1}\right]
\end{multline}
where $Y_i^{j-1}$ is the time extension of $Y$ from time instant $i$ to $j-1$. By adding $Y_i^{j-1}$ to the conditions of the second entropy, (\ref{eq})  will be increased to
\begin{equation}
\sum_{j=1}^n\left[\frac{h(Y_1(j),\ldots,Y_i(j)|T_{i,j},H(j))}{i}
-\frac{h(Y_1(j),\ldots,Y_{i-1}(j)|T_{i,j},H(j))}{i-1}\right]
\end{equation}
where $T_{i,j}=(W_1,\ldots,W_{i-1},Y_1^{j-1},\ldots,Y_i^{j-1},H^{j-1})$.
Therefore, we can write
\begin{align}\label{e30}
\sum_{i=1}^K\frac{nR_i}{i}  \leq &\overbrace{h(Y_1^n|H^n)}^{\leq n\log P} + \nonumber \\
&\sum_{i=2}^K\sum_{j=1}^n\left[\frac{h(Y_1(j),\ldots,Y_i(j)|T_{i,j},H(j))}{i}-
 \frac{h(Y_1(j),\ldots,Y_{i-1}(j)|T_{i,j},H(j))}{i-1}\right] + nO(\log P).
\end{align}
Before going further, the following lemma is needed.

\emph{Lemma 1}. Let $\Gamma_n=\{Y_1,Y_2,\ldots,Y_n\}$ be a set of $n(\geq2)$ arbitrary random variables and $\Omega_i^{j}(\Gamma_n)$ be a sliding window of size $j$ over $\Gamma_n$ i.e.,
\[\Omega_i^{j}(\Gamma_n) = Y_{(i-1)_n+1},Y_{(i)_n+1},\ldots,Y_{(i+j-2)_n+1}.\]
where $(.)_n$ defines the modulo $n$ operation. Then,
\begin{equation}\label{e24}
  (n-1)h(Y_1,Y_2,\ldots,Y_n|A)\leq \sum_{i=1}^{n}h(\Omega_i^{n-1}(\Gamma_n)|A).
\end{equation}
where $A$ is an arbitrary condition.

\begin{proof}
We prove the above by induction. It is obvious that (\ref{e24}) holds for $n=2$ (i.e., $h(Y_1,Y_2|A)\leq h(Y_1|A)+h(Y_2|A)$). Now, considering that (\ref{e24}) is valid for $n$, we show that it also holds for $n+1$. Replacing $n$ with $n+1$, we have
\begin{align}
    nh(Y_1,\ldots,Y_{n+1}|A)&= h(Y_1,\ldots,Y_{n+1}|A)+(n-1)h(Y_1,Y_2,\ldots,Y_{n-1},Z|A)\label{e26}\\
  &\leq h(Y_1,\ldots,Y_{n+1}|A)+ h(\Omega_1^{n-1}(\Psi_n)|A)+\sum_{i=2}^{n}h(\Omega_i^{n-1}(\Psi_n)|A)\label{e27}\\
  &= h(Y_1,\ldots,Y_{n+1}|A)+ h(Y_1,\ldots,Y_{n-1}|A)+\sum_{i=2}^{n}h(\Omega_i^{n}(\Gamma_{n+1})|A)\label{e28}
\end{align}
\begin{align}
&= h(Y_1,\ldots,Y_{n}|A)+h(Y_{n+1}|Y_1,\ldots,Y_n,A)+ h(Y_1,\ldots,Y_{n-1}|A)+\sum_{i=2}^{n}h(\Omega_i^{n}(\Gamma_{n+1})|A)\\
    &\leq h(Y_1,\ldots,Y_{n}|A)+h(Y_{n+1}|Y_1,\ldots,Y_{n-1},A)+ h(Y_1,\ldots,Y_{n-1}|A)+\sum_{i=2}^{n}h(\Omega_i^{n}(\Gamma_{n+1})|A)\label{e29}\\
  &= h(Y_1,\ldots,Y_{n}|A)+ h(Y_{n+1},Y_1,\ldots,Y_{n-1}|A)+\sum_{i=2}^{n}h(\Omega_i^{n}(\Gamma_{n+1})|A)\\
  &= \sum_{i=1}^{n+1}h(\Omega_i^{n}(\Gamma_{n+1})|A)
\end{align}
where in (\ref{e26}), $Z=(Y_n,Y_{n+1})$, in (\ref{e27}), $\Psi_n=\{Y_1,Y_2,\ldots,Y_{n-1},Z\}$ and the assumption of (\ref{e24}) being valid for $n$ is used, in (\ref{e28}), we have used the fact that $\Omega_i^{n-1}(\Psi_n)=\Omega_i^{n}(\Gamma_{n+1})$ (for $2\leq i \leq n$) and finally, (\ref{e29}) is due to the fact that conditioning reduces differential entropies. \qedhere
\end{proof}
Now, each term in the summation of (\ref{e30}) can be rewritten as
\begin{equation}\label{e31}
  \frac{(i-1)h(Y_1(j),\ldots,Y_i(j)|T_{i,j},H(j))-ih(Y_1(j),\ldots,Y_{i-1}(j)|T_{i,j},H(j))}{i(i-1)}
\end{equation}
and according to the previous lemma,
\begin{align}
  (\ref{e31})&\leq \frac{\sum_{r=1}^i\left[h(\Omega_r^{i-1}(\Gamma_i)|T_{i,j},H(j))-h(Y_1(j),\ldots,Y_{i-1}(j)|T_{i,j},H(j))\right]}{i(i-1)} \\
  &= \frac{\sum_{r=1}^{i-1}\left[h(Y_i(j)|E_r,T_{i,j},H(j))-h(Y_r(j)|E_r,T_{i,j},H(j))\right]}{i(i-1)}\label{e3e}
\end{align}
where $\Gamma_i=\{Y_1(j),Y_2(j),\ldots,Y_i(j)\}$, $E_r = \{Y_1(j),Y_2(j),...Y_{i-1}(j)\}-\{Y_r(j)\}$ and (\ref{e3e}) is from the chain rule of entropies. Before going further, the following lemma is needed.

\emph{lemma 2.} Consider a $K$-user MISO BC. At time instant $j$, two arbitrary users are selected with the received signals:
\begin{eqnarray}
  Y_m(j) &=& \textit{\textbf{h}}_m(j)^T\textit{\textbf{x}}(j)+w_m(j) \\
  Y_q(j) &=& \textit{\textbf{h}}_q(j)^T\textit{\textbf{x}}(j)+w_q(j).
\end{eqnarray}
Without loss of generality, we assume $m > q$. For simplicity, we assume that the communication is done in real dimensions where $\textit{\textbf{x}}\in R^{M\times 1}$ satisfying $E\left[\|\textit{\textbf{x}}\|^2\right]\leq P$, $\textit{\textbf{h}}_m$ and $\textit{\textbf{h}}_q$ have the distribution $N(\textbf{0},\textbf{I})$ and $w_m$ and $w_q$ have the distribution $N(0,1)$. When the CSIT of a user is either Perfect (P) or Not known (N), the following upper bound holds for the difference between entropies
\begin{equation}\label{e10}
 \lim_{P\to \infty} \frac{h(Y_m(j)|T)-h(Y_q(j)|T)}{\log P}\leq \left\{\begin{array}{cc} 1 & \mbox{ when CSIT of }\textit{\textbf{h}}_q(j) \mbox{ is } P \\ 0 & \mbox{ when CSIT of }\textit{\textbf{h}}_q(j) \mbox{ is } N  \end{array}\right.
\end{equation}
where $T$ is a condition such as the condition of entropies in (\ref{e3e}) or later in (\ref{eq1}). Interestingly, (\ref{e10}) is only a function of the CSIT of the second user. In other words, in the four possible cases of $PP,PN,NP$ and $NN$, the upper bound (not the exact value) for the pre-log factor of the difference is defined by the CSIT of the second user resulting in the same upper bound for the $PN$ or $NN$ case, and the same upper bound for the $PP$ or $NP$ case.

\begin{proof}
Based on the four possible states for the joint CSIT of $\textit{\textbf{h}}_m(j)$ and $\textit{\textbf{h}}_q(j)$, we have
\subsubsection{CSIT of $\textbf{\textit{h}}_m(j)$ is N or P and CSIT of $\textbf{\textit{h}}_q(j)$ is P}
\begin{equation}\label{e14}
  h(Y_m(j)|T)-h(Y_q(j)|T) \leq \underbrace{h(Y_m(j)|T)}_{\leq \log P}-\underbrace{h(Y_q(j)|T,W_1,\ldots,W_K)}_{O(\log P)}
\end{equation}
Since $\textbf{\textit{h}}_q(j)$ is known, a Gaussian input with the conditional covariance matrix of $\Sigma_{x|T}=P\textbf{\textit{u}}_q^{\perp}{\textbf{\textit{u}}_q^{\perp}}^H$ achieves the upper bound, where $\textbf{\textit{u}}_q^{\perp}$ is a unit vector in the direction orthogonal to $\textbf{\textit{h}}_q(j)$.
\subsubsection{CSIT of $\textbf{\textit{h}}_m(j)$ is N and CSIT of $\textbf{\textit{h}}_q(j)$ is N}
In this case both $Y_m(j)$ and $Y_q(j)$ are statistically equivalent (i.e., having the same probability density functions, and subsequently, the same entropies.) Therefore,
\begin{equation}
  h(Y_m(j)|T)-h(Y_q(j)|T)=0
\end{equation}
\subsubsection{CSIT of $\textbf{\textit{h}}_m(j)$ is P and CSIT of $\textbf{\textit{h}}_q(j)$ is N}
This is a rather more complicated scenario and we defer the proof to Appendix \ref{s6}.
\qedhere
\end{proof}
From (\ref{e30}) and (\ref{e3e}), we have
\begin{align}
  \sum_{i=1}^K \frac{nR_i}{i} &\leq n\log P + \sum_{i=2}^K\sum_{j=1}^n\sum_{r=1}^{i-1}\frac{h(Y_i(j)|E_r,T_{i,j},H(j))-h(Y_r(j)|E_r,T_{i,j},H(j))}{i(i-1)} +nO(\log P)\\
&= n\log P + \sum_{i=2}^K\sum_{r=1}^{i-1}\frac{\sum_{j=1}^n\left[h(Y_i(j)|E_r,T_{i,j},H(j))-h(Y_r(j)|E_r,T_{i,j},H(j))\right]}{i(i-1)}+nO(\log P)\label{e47e} \\
&\leq n\log P + \sum_{i=2}^K\sum_{r=1}^{i-1}\frac{n\lambda_P^r}{i(i-1)}\log P +nO(\log P) \label{e4e}
\end{align}
where (\ref{e4e}) is from the application of lemma 2 and the fact that $n$ is sufficiently large. Therefore,
\begin{equation}
  \sum_{i=1}^K\frac{d_i}{i}\leq 1 + \sum_{i=2}^K\frac{\sum_{r=1}^{i-1}\lambda_P^r}{i(i-1)}.
\end{equation}
It is obvious that the same approach can be applied to any other permutations on $(1,2,\ldots,K)$. In addition to the mentioned proof, an alternative converse is provided in Appendix \ref{s4}.
\subsection{Proof of $\sum_{i=1}^K d_i \leq 1 + \sum_{i=1}^{K-1}(\lambda_P^{\psi_{\pi^K}(i)} + \lambda_D^{\psi_{\pi^K}(i)})$}
We enhance the channel in two ways:
\begin{enumerate}
  \item Like the approach in \cite{Tandon}, whenever there is delayed CSIT ($D$), we assume that it is perfect instantaneous CSIT ($P$), but we keep the probability of delayed CSIT. In other words, the CSIT of user $i$ is perfect with probability $\lambda_P^i+\lambda_D^i$ and unknown otherwise.
  \item We give the message of user $i$ ($W_i$) to users $i+1,i+2,\ldots,K$.
\end{enumerate}
  Therefore,
  \begin{equation}\label{e3}
    nR_i\leq I(W_i;Y_i^n|W_1,\ldots,W_{i-1},H^n).
  \end{equation}
   By summing (\ref{e3}) over users and writing the mutual informations in terms of differential entropies,
   \begin{equation}
     \sum_{i=1}^KnR_i\leq \overbrace{h(Y_1^n|H^n)}^{\leq n\log P}+ \sum_{i=2}^K\left[h(Y_i^n|W_1,\ldots,W_{i-1},H^n)-
      h(Y_{i-1}^n|W_1,\ldots,W_{i-1},H^n)\right ] + nO(\log P).
  \end{equation}
  The term in the summation could be written as
  \begin{multline}\label{e33}
   h(Y_i^n|W_1,\ldots,W_{i-1},H^n)-
      h(Y_{i-1}^n|W_1,\ldots,W_{i-1},H^n)\\
   =\sum_{j=1}^n\left[h(Y_i(j)|T_{i,j},H(j))-h(Y_{i-1}(j)|T_{i,j},H(j))\right]
  \end{multline}
  where
  \begin{equation*}
    T_{i,j}=\left (W_1,\ldots,W_{i-1},H^{j-1},Y_{i-1}(1),\ldots,Y_{i-1}(j-1),Y_i(j+1),\ldots,Y_i(n)\right).
  \end{equation*}
 The proof of (\ref{e33}) is provided in Appendix \ref{s5} .Therefore,
 \begin{equation}\label{eq1}
   \sum_{i=1}^KnR_i\leq n\log P + \sum_{i=2}^K\sum_{j=1}^n\left[h(Y_i(j)|T_{i,j},H(j))-h(Y_{i-1}(j)|T_{i,j},H(j))\right]
 \end{equation}
and finally, by applying the results of lemma 2 to (\ref{eq1}), we have
\begin{equation}\label{e373}
   \sum_{i=1}^K d_i \leq 1+\sum_{i=2}^K(\lambda_P^{i-1}+\lambda_D^{i-1})=1+\sum_{i=1}^{K-1}(\lambda_P^{i}+\lambda_D^{i}).
\end{equation}
Let $\pi^K(.)$ be an arbitrary permutation of size $K$ on $(1,\ldots,K)$. Applying the same reasoning, we have
\begin{equation}\label{e37}
   \sum_{i=1}^K d_i \leq 1+\sum_{i=1}^{K-1}(\lambda_P^{\pi^K(i)}+\lambda_D^{\pi^K(i)})\ \ \ \ \forall \pi^K(.)
\end{equation}
(\ref{e37}) results in $K$ inequalities all having the same left hand side. Therefore,
\begin{equation}\label{ee37}
   \sum_{i=1}^K d_i \leq 1+\min_{\pi^K(.)}{\sum_{i=1}^{K-1}(\lambda_P^{\pi^K(i)}+\lambda_D^{\pi^K(i)})}
\end{equation}
and it is obvious that $\psi_{\pi^K}(.)$ will minimize (\ref{ee37}) if and only if it satisfies (\ref{const}) (for $j=K$.)
\section{Achievability}\label{s55}
In this section, we consider the achievability of the symmetric case. The outer bound in theorem consists of $2^K-1+\sum_{j=2}^Kj!\left(\begin{array}{c}K\\j\end{array}\right)$ inequalities. For $K=2$, it is observed that the outer bound (even with $M>2$) will be the same as \cite{Tandon} where it was shown to be achievable, regardless of the pattern of CSIT. For $K\geq 3$, we show that given the marginal probabilities of CSIT, there exists at least one CSIT pattern that achieves the outer bound in some scenarios. We investigate the following two cases:
\subsection{$\lambda_D = 0$}
\begin{figure}[t]
  \centering
  \includegraphics[width=10cm]{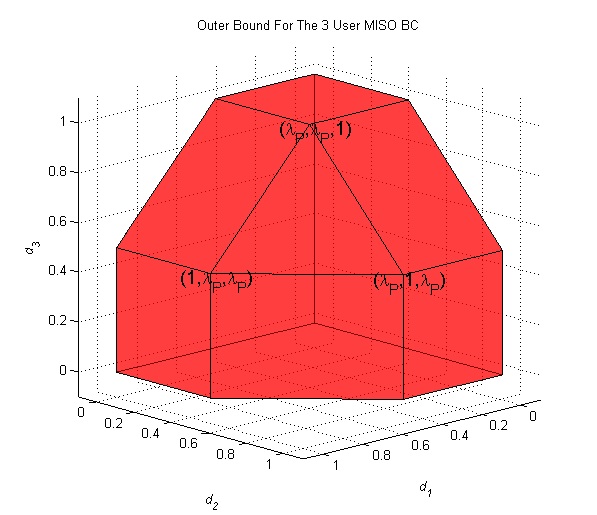}\\
  \caption{Region in case A for 3 user BC}\label{fig1}
\end{figure}

\begin{figure}[t]
  \centering
  \includegraphics[width=8cm]{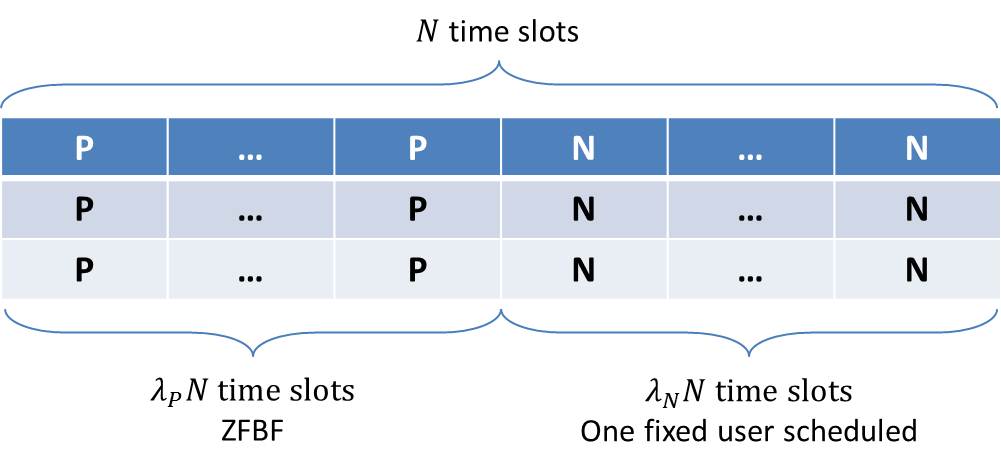}\\
  \caption{Achievable scheme in case A for 3 user BC}\label{fig2}
\end{figure}

In this case, $2^K-1$ inequalities ($2^K-K-1$ inequalities having $\sum_i d_i$ (summation with equal weights) in the left-hand side and $K$ single user inequalities) are active and the remaining $\sum_{j=2}^Kj!\left(\begin{array}{c}K\\j\end{array}\right)$ inequalities become inactive. The reason can be easily verified from the inequalities, however, a simpler intuitive way is to consider that those $\sum_{j=2}^Kj!\left(\begin{array}{c}K\\j\end{array}\right)$ inequalities are derived from making the channel degraded and when there is no delayed CSIT, this degradation results in loose bounds. Equivalently, when there is no delayed CSIT, those inequalities derived from the degraded broadcast channel are inactive. In this case, the region is defined by $2^K-1$ hyperplanes in $R_+^K$ and has the following K corner points:
\[ (1,\lambda_P,\ldots,\lambda_P),(\lambda_P,1,\lambda_P,\ldots,\lambda_P),\ldots,(\lambda_P,\ldots,\lambda_P,1)\]
The corner points have the unique characteristic that the whole region can be constructed by time sharing between them. Therefore, the achievability of these points is equivalent to the achievability of the whole region.
Figure \ref{fig1} shows the region for the 3 user broadcast channel. The corner points are simply achieved by the scheme shown in figure \ref{fig2}.
The scheme has $N$ time slots and consists of two parts: in the first $\lambda_PN$ time slots, zero forcing beamforming (ZFBF) is carried out where each user receives one interference-free symbol. In the remaining $\lambda_NN$ time slots, only one particular user (depending on the corner point of interest) is scheduled.

\subsection{$\lambda_N\leq \frac{\lambda_D}{\sum_{j=2}^K\frac{1}{j}}$}
Before going further, we need the following simple lemma.

\textit{Lemma 3}. The minimum probability of delayed CSIT for sending order-$j$ symbols in the $K$-user MAT is
\begin{equation}\label{e9}
  \lambda_D^{min}(K,j)=1-\frac{K-j+1}{K\sum_{i=j}^K\frac{1}{i}}.
\end{equation}
Substituting $j=1$ in (\ref{e9}), we get the minimum $\lambda_D$ for order-$1$ symbols as
\begin{equation}
  \lambda_D^{min}(K)=1-\frac{1}{\sum_{i=j}^K\frac{1}{i}}.
\end{equation}
\begin{proof} From \cite{MAT}, the MAT algorithm is based on a concatenation of $K$ phases. As shown below, phase $j$ takes $(K-j+1)\left(\begin{array}{c}K\\j\end{array}\right)$ order-$j$ messages as its input, takes $\left(\begin{array}{c}K\\j\end{array}\right)$  time slots and produces $j\left(\begin{array}{c}K\\j+1\end{array}\right)$ order-$j+1$ messages as its output, as illustrated below
\[(K-j+1)\left(\begin{array}{c}K\\j\end{array}\right) \mbox{order-}j\to \underbrace{{\mbox{Phase }j}}_{\left(\begin{array}{c}K\\j\end{array}\right)\mbox{time slots}}\to j\left(\begin{array}{c}K\\j+1\end{array}\right)\mbox{order-}j+1.\]
In each time slot of phase $j$, the transmitter sends a random linear combination of the $(K-j+1)$ symbols to a subset $S$ of receivers , $|S|=j$. Sending the overheard interferences from the remaining $(K-j)$ receivers to receivers in subset S enables them to successfully decode their $(K-j+1)$ symbols by constructing a set of $(K-j+1)$ linearly independent equations. Therefore, the transmitter needs to know the channel of only $(K-j)$ receivers. In other words, at each time slot of phase $j$, the feedback of $(K-j)$ CSI is enough.
In the MAT algorithm the number of output symbols that phase $j$ produces should match the number of input symbols of phase $j+1$. The ratio between the input of phase $j+1$ and output of phase $j$ is:
\[\frac{(K-j)\left(\begin{array}{c}K\\j+1\end{array}\right)}{j\left(\begin{array}{c}K\\j+1\end{array}\right)}=\frac{(K-j)}{j}.\]
This means that $(K-j)$ repetition of phase $j$ will produce the inputs needed by $j$ repetition of phase $j+1$. In general, in order to have an integer number for repetitions, we multiply phase $1$ by $K!$ (i.e., repeat it $K!$ times), phase $2$ by $\frac{K!}{(K-1)}$, and so on. Therefore, phase $j$ will be repeated $((j-1)!(K-j)!)K$ times which takes $((j-1)!(K-j)!)K\left(\begin{array}{c}K\\j\end{array}\right)$ time slots. Since $(K-j)$ feedbacks from each time slot is sufficient, the number of feedbacks will be $((j-1)!(K-j)!)K\left(\begin{array}{c}K\\j\end{array}\right)(K-j)$. For a successive decoding or order-$j$ symbols, all the higher order symbols must be decoded successfully. Therefore, instead of having delayed CSIT at all time instants from all users, the minimum probability of delayed CSIT is the number of feedbacks from phase $j$ to $K$ divided by the whole number of time slots multiplied by the number of users,
\[\lambda_D^{min}(K,j)=\frac{\sum_{i=j}^K(i-1)!(K-i)!K\left(\begin{array}{c}K\\i\end{array}\right)(K-i)}
{\sum_{i=j}^K(i-1)!(K-i)!K\left(\begin{array}{c}K\\i\end{array}\right)K}=1-\frac{K-j+1}{K\sum_{i=j}^K\frac{1}{i}}.\] \qedhere
\end{proof}
\begin{figure}[t]
  \centering
  \includegraphics[width=10cm]{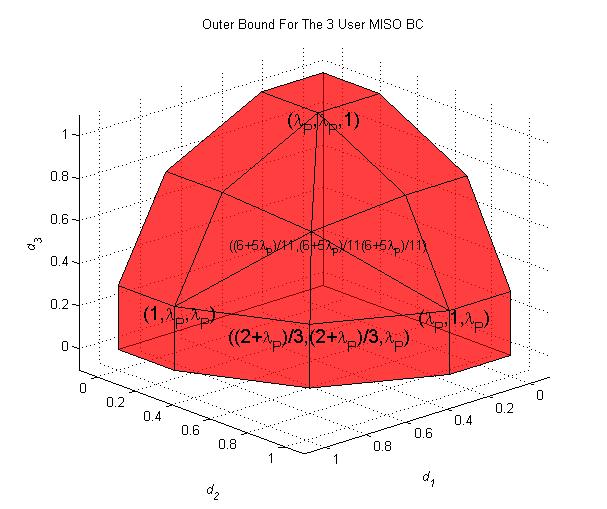}\\
  \caption{Region in case B for 3 user BC}\label{fig3}
\end{figure}
In this case (i.e., $\lambda_N\leq \frac{\lambda_D}{\sum_{j=2}^K\frac{1}{j}}$), the $2^K-K-1$ inequalities having $\sum_i d_i$ (summation with equal weights) in the left-hand side become inactive and the remaining $\sum_{j=1}^Kj!\left(\begin{array}{c}K\\j\end{array}\right)$  inequalities are active which construct $\sum_{j=1}^Kj!\left(\begin{array}{c}K\\j\end{array}\right)$  hyperplanes in $R_+^K$. This region has the following $2^K-1$ corner points
\begin{itemize}
  \item $\left(\begin{array}{c}K\\1\end{array}\right) \mbox{corner points in the form } (1,\lambda_P,\ldots,\lambda_P),(\lambda_P,1,\lambda_P,\ldots,\lambda_P),\ldots,(\lambda_P,\ldots,\lambda_P,1) $
  \item $\left(\begin{array}{c}K\\2\end{array}\right) \mbox{corner points in the form } (\frac{2+\lambda_P}{3},\frac{2+\lambda_P}{3},\lambda_P,\ldots,\lambda_P),(\frac{2+\lambda_P}{3},\lambda_P,\frac{2+\lambda_P}{3},\lambda_P,\ldots,\lambda_P),\ldots $
  \item $\left(\begin{array}{c}K\\3\end{array}\right) \mbox{corner points in the form } (\frac{6+5\lambda_P}{11},\frac{6+5\lambda_P}{11},\frac{6+5\lambda_P}{11},\lambda_P,\ldots,\lambda_P),\ldots $
  \item \ \ $\ldots, \mbox{and finally,}\left(\begin{array}{c}K\\K\end{array}\right)\mbox{corner points in the form }(\frac{1+\lambda_P\sum_{i=2}^K\frac{1}{i}}{\sum_{i=1}^K\frac{1}{i}},\frac{1+\lambda_P\sum_{i=2}^K\frac{1}{i}}{\sum_{i=1}^K\frac{1}{i}},\ldots,\frac{1+\lambda_P\sum_{i=2}^K\frac{1}{i}}{\sum_{i=1}^K\frac{1}{i}})$
\end{itemize}
The region for the 3 user broadcast channel and the achievable scheme are shown in figure \ref{fig3} and figure \ref{fig4}, respectively. The scheme is based on a concatenation of ZFBF and MAT as follows.
For the first $K$ corner points listed above, the achievability scheme is the same as that in the previous section (i.e., ZFBF + fixed user scheduling). For the $\left(\begin{array}{c}K\\j\end{array}\right)$ ($j\geq 2$) corner points in the form $(\frac{1+\lambda_P\sum_{i=2}^j\frac{1}{i}}{\sum_{i=1}^j\frac{1}{i}},\frac{1+\lambda_P\sum_{i=2}^j\frac{1}{i}}{\sum_{i=1}^j\frac{1}{i}},\ldots,\lambda_P,\ldots,\lambda_P)$, we write
\begin{equation}\
  \lambda_P=\frac{M_1}{N_1}, \lambda_D=\frac{M_2}{N_2}, \lambda_D^{min}(j)=\frac{m}{n}
\end{equation}
where $\lambda_D^{min}(j)$ is the minimum probability of delayed CSIT for sending order-1 symbols in the $j$-user MAT. $m,n,M_i$  and $N_i$ ($i=1,2$) are integers. Making a common denominator between $\lambda_P$ and $\lambda_D$ we have
\begin{equation}
 \lambda_P=\frac{nM_1N_2}{nN_1N_2}, \lambda_D=\frac{nN_1M_2}{nN_1N_2}.
\end{equation}
\begin{figure}
  \centering
  \includegraphics[width=8cm]{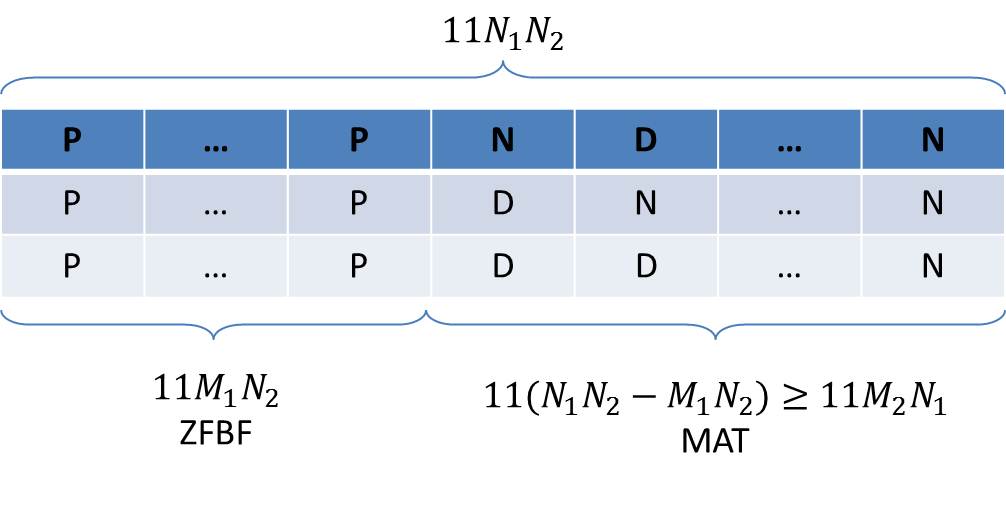}\\
  \caption{Achievable scheme in case B for 3 user BC}\label{fig4}
\end{figure}
We construct $nN_1N_2$ time slots where the CSIT of each user can be Perfect (P) or Delayed (D) in $nM_1N_2$ or $nN_1M_2$ time slots, respectively. In the first $nM_1N_2$ time slots, ZFBF is carried out. In the remaining $n(N_1N_2-M_1N_2)$ time slots, $j$-user MAT algorithm is done.
At each time slot of the ZFBF part, 1 interference-free symbol is received by each user and in the MAT part, $\frac{n(N_1N_2-M_1N_2)}{1+\frac{1}{2}+\cdots+\frac{1}{j}}$ symbols are sent to each of the users in subset $S$ (with $|S|=j$) where $S$ depends on the corner point of interest. In order to do the MAT algorithm in the second part, the minimum probability of delayed CSIT should be met
\begin{equation}
  nN_1M_2 \geq \lambda_D^{min}(j)n(N_1N_2-M_1N_2)
\end{equation}
Dividing both sides by $nN_1N_2$,
\begin{equation}
  \lambda_D \geq \lambda_D^{min}(j)(1-\lambda_P)=\lambda_D^{min}(j)(\lambda_D+\lambda_N)
\end{equation}
which results in
\begin{equation}
 \lambda_N \leq \frac{\lambda_D}{\sum_{i=2}^j\frac{1}{i}}.
\end{equation}
Since it should be valid for all $j$, we have
\begin{equation}
 \lambda_N \leq \frac{\lambda_D}{\sum_{i=2}^K\frac{1}{i}}
\end{equation}
which is the condition assumed in this case. For the case when $\lambda_N > \frac{\lambda_D}{\sum_{i=2}^K\frac{1}{i}}$, finding a general achievable scheme remains an open problem.

\begin{figure}
  \centering
  \includegraphics[width=6cm]{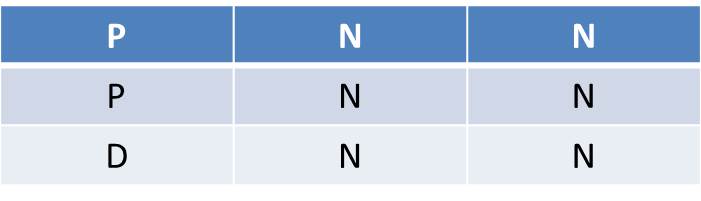}\\
  \caption{The CSIT pattern of the example with $\lambda_P^1=\lambda_P^2=\lambda_D^3=\frac{1}{3} \mbox{ and } \lambda_N^i=\frac{2}{3}, i=1,2,3.$}\label{fig5}
\end{figure}

In case B, the $2^K-K-1$ inequalities having $\sum_i d_i$ (summation with equal weights) in the left-hand side were inactive. From section $B$ in the proof of theorem, these inequalities were derived by enhancing the channel in a way that whenever there is delayed CSIT we replace it with Perfect CSIT as in \cite{Tandon} for the two user case. Due to this enhancement, a question may be raised whether these inequalities are always loose or not when $\lambda_D \neq0$ and $K>2$. The following example shows that these inequalities are not always loose. Consider the CSIT pattern in figure \ref{fig5} where $\lambda_P=\frac{2}{9},\lambda_D=\frac{1}{9} \mbox{ and } \lambda_N=\frac{6}{9}$. From the outer bound in theorem, it is observed that the sum DoF must be lower than or equal to $\frac{5}{3}$. Actually the sum DoF of $\frac{5}{3}$ is optimal and achievable as shown in the sequel.

The transmitted signal at time slot $1$ is
\begin{equation*}
 \textit{\textbf{ x}}(1)=\textbf{P}_1\textbf{\textit{u}}_1 + \textbf{P}_2\textbf{\textit{u}}_2 + u_3\textit{\textbf{v}}
\end{equation*}
where $\textbf{\textit{u}}_1$ and $\textbf{\textit{u}}_2$ are the (2 by 1) private message vectors for user 1 and 2, respectively. $u_3$ is the private (scalar) message for user 3 and $\textit{\textbf{v}}$ is a (3 by 1) vector orthogonal to both $\textit{\textbf{h}}_1$ and $\textit{\textbf{h}}_2$. $\textbf{P}_1$ and $\textbf{P}_2$ are the (3 by 2) precoding matrices with the following property:
\begin{equation*}
  \textit{\textbf{h}}_2^H\textbf{P}_1=\textit{\textbf{h}}_1^H\textbf{P}_2=\textit{\textbf{0}}_{1\times 2}
\end{equation*}
At time slot 1, the received signals at the receivers are
\begin{align*}
  y_1(1) &= L_1(\textbf{\textit{u}}_1) \\
  y_2(1) &= L_2(\textbf{\textit{u}}_2) \\
  y_3(1) &= \tilde{L}_1(\textbf{\textit{u}}_1)+\tilde{L}_2(\textbf{\textit{u}}_2)+(\textit{\textbf{h}}_3^H\textit{\textbf{v}})u_3.
\end{align*}
where $L_1(\textbf{\textit{u}}_1)=\textit{\textbf{h}}_1^H\textbf{P}_1\textbf{\textit{u}}_1,\  L_2(\textbf{\textit{u}}_2)=\textit{\textbf{h}}_2^H\textbf{P}_2\textbf{\textit{u}}_2,\  \tilde{L}_1(\textbf{\textit{u}}_1)=\textit{\textbf{h}}_3^H\textbf{P}_1\textbf{\textit{u}}_1$ and $\tilde{L}_2(\textbf{\textit{u}}_2)=\textit{\textbf{h}}_3^H\textbf{P}_2\textbf{\textit{u}}_2$. The transmitter sends $\tilde{L}_1(\textbf{\textit{u}}_1)$ and $\tilde{L}_2(\textbf{\textit{u}}_2)$ at time slots $2$ and $3$, respectively. Having $L_1$ and $\tilde{L}_1$, receiver $1$ can decode its two private messages, since $\textit{\textbf{h}}_1$ and $\textit{\textbf{h}}_3$ are statistically and hence, linearly independent almost surely. The same applies to receiver $2$, and receiver $3$ can decode its private message after eliminating the interference terms $\tilde{L}_1$ and $\tilde{L}_2$. Thus, the inequality ($d_1 +d_2 +d_3 \leq \frac{5}{3}$) is tight in this case.

\begin{figure}
  \centering
  \includegraphics[width=10 cm]{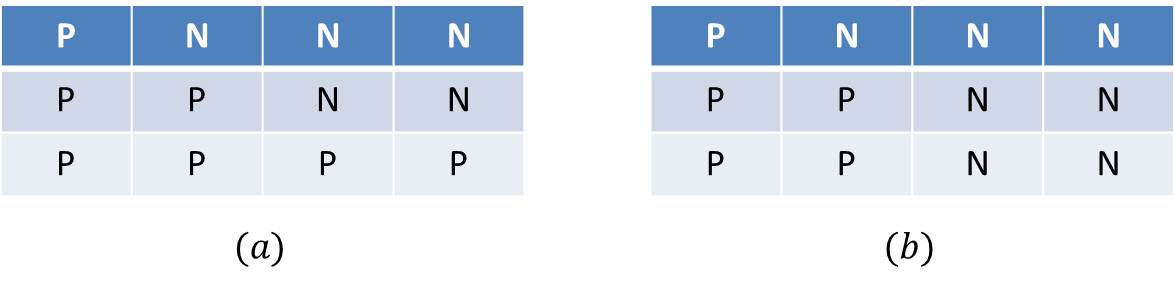}\\
  \caption{Two CSIT patterns. $(a)\  \lambda_P^1=\frac{1}{4},\lambda_P^2=\frac{1}{2} \mbox{ and } \lambda_P^3=1\ (b)\  \lambda_P^1=\frac{1}{4} \mbox{ and } \lambda_P^2=\lambda_P^3=\frac{1}{2}$}\label{fig6}
\end{figure}

For the asymmetric scenario with no Delayed (D) CSIT, it is interesting to note that the outer bound in theorem does not depend on the maximum probability of Perfect CSIT. Since there is no delayed CSIT, those inequalities obtained from making the channel degraded are inactive and only the inequality having the form of summation with equal weights becomes active. According to (\ref{theorem}) and (\ref{const}), the user with the highest probability of perfect CSIT is excluded from the right-hand side of the inequality. For example, according to the theorem, sum DoF of the two CSIT patterns shown in figure \ref{fig6} has the upper bound of $\frac{7}{4}$ and since it is achievable in both patterns, it is optimal.
\section{Dependency of the dof region on the csit pattern}\label{sh}
\begin{figure}[t]
  \centering
  \includegraphics[width=8cm]{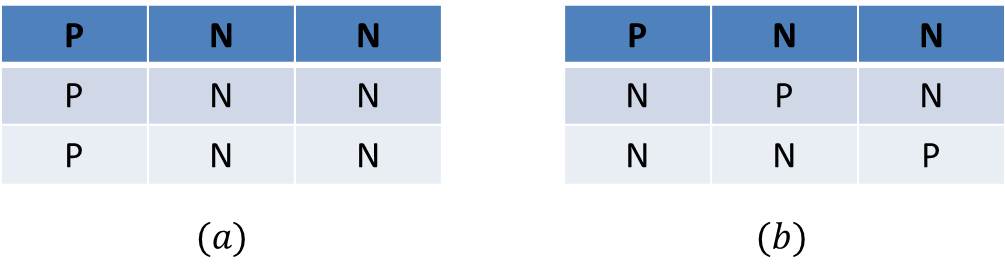}\\
  \caption{Two symmetric CSIT patterns having the same marginal probabilities (i.e., $\lambda_P=1-\lambda_N=\frac{1}{3}$.)}\label{fig100}
\end{figure}

In the previous sections, the focus was on the outer bounds and their achievabilities given only the marginal probabilities. Here, we show that two different CSIT patterns, though having the same marginal probabilities, do not necessarily have the same DoF region. Consider the two simple symmetric CSIT patterns shown in figure \ref{fig100}. According to the theorem, the DoF region of both has an outer bound as shown in figure \ref{fig1} with the corner points $(1,\frac{1}{3},\frac{1}{3}),(\frac{1}{3},1,\frac{1}{3})$ and $(\frac{1}{3},\frac{1}{3},1)$. It is obvious that the corner points are achievable for pattern $(a)$, and in what follows we show that they are not achievable for pattern $(b)$. In other words, the DoF region of pattern $(b)$ is inside that of pattern $(a)$. We write,
\begin{align}
  nR_1 &\leq I(W_1;Y_1^n|H^n) \label{eq14}\\
  nR_1 &\leq I(W_1;Y_1^n|H^n,W_2)\label{eq15}
\end{align}
where in (\ref{eq15}), we use the fact that $W_1$ and $W_2$ are independent. Adding (\ref{eq14}) and (\ref{eq15}) results in
\begin{equation}\label{eq16}
  2nR_1 \leq I(W_1;Y_1^n|H^n)+ I(W_1;Y_1^n|H^n,W_2).
\end{equation}
By doing the same for $R_2$, we have
\begin{equation}\label{eq17}
  2nR_2 \leq I(W_2;Y_2^n|H^n)+ I(W_2;Y_2^n|H^n,W_1).
\end{equation}
Finally, the rate of user 3 is written as
\begin{equation}\label{eq18}
  nR_3 \leq I(W_3;Y_3^n|H^n,W_1,W_2).
\end{equation}
By adding (\ref{eq16}),(\ref{eq17}) and (\ref{eq18}) and writing them in terms of differential entropies, we get
\begin{align}
  2nR_1+2nR_2+nR_3 &\leq \underbrace{h(Y_1^n|H^n)}_{\leq n\log P}+\underbrace{h(Y_2^n|H^n)}_{\leq n\log P}  \\
  &\ \ \underbrace{+h(Y_2^n|H^n,W_1)-h(Y_1^n|H^n,W_1)}_{\leq \frac{n}{3}\log P}\underbrace{+h(Y_1^n|H^n,W_2)-h(Y_2^n|H^n,W_2)}_{\leq \frac{n}{3}\log P}\label{a0}\\
  &\ \ + h(Y_3^n|H^n,W_1,W_2)\underbrace{-h(Y_1^n|H^n,W_1,W_2)-h(Y_2^n|H^n,W_1,W_2)}_{\leq -h(Y_1^n,Y_2^n|H^n,W_1,W_2)} \label{53}\\
  &\leq \frac{8n}{3}\log P + h(Y_3^n|H^n,W_1,W_2)-h(Y_1^n,Y_2^n|H^n,W_1,W_2)\label{a1}\\
  &= \frac{8n}{3}\log P +\underbrace{h(Y_3^n|H^n,W_1,W_2)-h(Y_{2,PNN}^n,Y_{1,NPN}^n,Y_{1,NNP}^n|H^n,W_1,W_2)}_{O(\log P)}\label{a2}\\
  &\ \ \underbrace{-h(Y_{1,PNN}^n,Y_{2,NPN}^n,Y_{2,NNP}^n|H^n,W_1,W_2,Y_{2,PNN}^n,Y_{1,NPN}^n,Y_{1,NNP}^n)}_{\leq -h(Y_{1,PNN}^n,Y_{2,NPN}^n,Y_{2,NNP}^n|H^n,W_1,W_2,Y_{2,PNN}^n,Y_{1,NPN}^n,Y_{1,NNP}^n,W_3)\sim O(\log P)}\label{a3}\\
  &\leq \frac{8n}{3}\log P
\end{align}
where in (\ref{a0}), the difference terms are first written as a time summation of instantaneous differences, as in (\ref{e33}). Then, lemma 2 of section \ref{ss4} is applied to the differences resulting in the values written under braces. We have split the observation of users 2 and 3 in terms of the joint CSIT, i.e., $Y_1^n=(Y_{1,PNN}^n,Y_{1,NPN}^n,Y_{1,NNP}^n)$ and $Y_2^n=(Y_{2,PNN}^n,Y_{2,NPN}^n,Y_{2,NNP}^n)$. Again, in (\ref{a2}), the difference terms are first written as a time summation of instantaneous differences and by the application of lemma 2 we get the upperbound shown under the braces. Specifically, (\ref{a2}) is due to the fact that there is at least one unknown CSIT (N) in the joint observations $Y_1^n,Y_2^n$ (see rows 1 and 2 of the CSIT pattern shown in figure \ref{fig100} $(b)$.) Finally, (\ref{a3}) is due to the fact that conditioning reduces the entropy and knowledge of all the messages and the channels enable us to reconstruct each observation within noise distortion. Therefore, for pattern $(b)$, the following inequalities hold which make its DoF region inside that of pattern $(a)$.
\begin{align}\label{ew1}
  2d_1+2d_2+d_3 &\leq \frac{8}{3} \\
  2d_1+d_2+2d_3 &\leq \frac{8}{3} \\
  d_1+2d_2+2d_3 &\leq \frac{8}{3}.
\end{align}
Motivated by this simple example, we can have the following set of inequalities for the 3-user MISO BC
%\begin{align}
%  2nR_1+2nR_2+nR_3 &\leq \underbrace{h(Y_1^n|H^n)}_{\leq n\log P}+\underbrace{h(Y_2^n|H^n)}_{\leq n\log P}  \\
%  &\ \underbrace{+h(Y_2^n|H^n,W_1)-h(Y_1^n|H^n,W_1)}_{\leq n(\lambda_P^1+\lambda_D^1)\log P}\underbrace{+h(Y_1^n|H^n,W_2)-h(Y_2^n|H^n,W_2)}_{\leq n(\lambda_P^2+\lambda_D^2)\log P}\\
%  &\ \underbrace{+ h(Y_3^n|H^n,W_1,W_2)\underbrace{-h(Y_1^n|H^n,W_1,W_2)-h(Y_2^n|H^n,W_1,W_2)}_{\leq -h(Y_1^n,Y_2^n|H^n,W_1,W_2)}}_{\leq n(\lambda_{PP-}+\lambda_{PD-}+\lambda_{DP-}+\lambda_{DD-})\log P}
%\end{align}
\begin{align}\label{ew1}
  2d_1+2d_2+d_3 &\leq 2+(\lambda_P^1+\lambda_D^1)+(\lambda_P^2+\lambda_D^2)+(\lambda_{PP-}+\lambda_{PD-}+\lambda_{DP-}+\lambda_{DD-}) \\
  2d_1+d_2+2d_3 &\leq 2+(\lambda_P^1+\lambda_D^1)+(\lambda_P^3+\lambda_D^3)+(\lambda_{P-P}+\lambda_{P-D}+\lambda_{D-P}+\lambda_{D-D}) \\
  d_1+2d_2+2d_3 &\leq 2+(\lambda_P^2+\lambda_D^2)+(\lambda_P^3+\lambda_D^3)+(\lambda_{-PP}+\lambda_{-PD}+\lambda_{-DP}+\lambda_{-DD})
\end{align}
where a dashed line in the above means that the CSIT of the corresponding user is not important (for example, $\lambda_{PD-}=\lambda_{PDP}+\lambda_{PDD}+\lambda_{PDN}$ which is a summation over all the possible values for the CSIT of user 3). The same approach could be easily extended to the $K$-user MISO BC which is omitted for brevity. It is obvious that none of the above inequalities can have its right-hand side written in terms of only marginal probabilities. Therefore, in contrast to the two user scenario, marginal probabilities of CSIT are not sufficient for defining the DoF region of the general $K$-user MISO BC, and having the same marginal probabilities does not guarantee the same DoF region.
\section{Conclusion}\label{s7}
Given the marginal probabilities of CSIT, an outer bound was derived for the DoF region of the $K$-user MISO BC with CSIT alternating among Perfect (P), Delayed (D) or Not known (N). This outer bound was shown to be achievable by specific CSIT patterns in certain regions. Through an example, we showed that in general, the DoF region of the $K$-user MISO BC (when $K\geq 3$) is a function of CSIT patterns or equivalently the $3^K$ state probabilities rather than the sole marginal probabilities. In contrast to the two user case, the following items may be the reason why the outer bounds may not always be achievable: 1) It should be a function of all joint probabilities. 2) Making the channel degraded when there are also P and N CSITs. 3) Enhancing the channel by treating D as P. Investigating this problem in more detail and the generalization of the results to the MIMO BC are the topics of our future work.
\appendices
\section{Proof of case (3) in lemma 2}\label{s6}
We use the following lemma, which is a simple generalization of corollary 6 in \cite{extremal}.

\emph{Lemma 4.} Let $\hat{Z}_1$ and $\hat{Z}_2$ be two Gaussian random variables, and let $\textit{\textbf{a}}$ and $\textit{\textbf{b}}$ be two deterministic vectors in $R^M$. Let $U$ be a random variable independent of $\hat{Z}_1$ and $\hat{Z}_2$. In the optimization problem
\begin{align}\label{e100}
  \max_{p(\textit{\textbf{x}}|u)} \ \ &h(\textit{\textbf{a}}^T\textit{\textbf{X}}+\hat{Z}_1|U) - \mu h(\textit{\textbf{b}}^T\textit{\textbf{X}}+\hat{Z}_2|U)\\
\mbox{subject to}\ \  &Cov(\textit{\textbf{X}}|U)\preceq \textbf{\textit{S}} \mbox{\ and\ } \textit{\textbf{X}}\to U\to (\hat{Z}_1,\hat{Z}_2)
\end{align}
for any $\mu\geq 1$ and positive semidefinite $\textbf{\textit{S}}$, a Gaussian $p(\textit{\textbf{x}}|u)$ with the same covariance matrix for each $u$ is an optimal solution.
\begin{proof} We restate theorem 8 in \cite{extremal}: Let $\textbf{\textit{Z}}_1$ and $\textbf{\textit{Z}}_2$ be two Gaussian vectors with strictly positive definite covariance matrices $\textbf{\textit{K}}_{Z_1}$ and $\textbf{\textit{K}}_{Z_2}$, respectively. Let $\mu\geq 1$ be a real number, $\textbf{\textit{S}}$ be a positive semidefinite matrix and $U$ be a random variable independent of $\textbf{\textit{Z}}_1$ and $\textbf{\textit{Z}}_2$. Consider the optimization problem
\begin{align}\label{e100}
  \max_{p(\textit{\textbf{x}}|u)} \ \ &h(\textit{\textbf{X}}+\textit{\textbf{Z}}_1|U) - \mu h(\textit{\textbf{X}}+\textit{\textbf{Z}}_2|U)\\
\mbox{subject to}\ \  &Cov(\textit{\textbf{X}}|U)\preceq \textbf{\textit{S}}
\end{align}
where the maximization is over all conditional distribution of $\textit{\textbf{X}}$ given
$U$ independent of $\textbf{\textit{Z}}_1$ and $\textbf{\textit{Z}}_2$. A Gaussian $p(\textit{\textbf{x}}|u)$ with the same covariance
matrix for each $u$ is an optimal solution of this optimization problem. The proof follows the same steps as in theorem 1 in \cite{extremal} with replacing the classical Entropy Power Inequality (EPI) by its conditional version \cite{Bergmans}.

Now, instead of (150) in Appendix D of \cite{extremal}, (\ref{e100}) can be used. Next, lemma 13 in \cite{extremal} is generalized to the $M$ dimensional conditional version as follows.

Let $\textbf{\textit{Z}} = (Z_1,Z_2,\ldots,Z_M)^t$ where $Z_1,Z_2,\ldots,Z_M$ are independent Gaussian variables with variances $\sigma_1^2,\sigma_2^2,\ldots,\sigma_M^2$, respectively. Let $U$ be a random variable independent of $\textbf{\textit{Z}}$. For any random vector $\textbf{\textit{X}} = (X_1,X_2,\ldots,X_M)^t$ with finite variances and given $U$ independent of $\textbf{\textit{Z}}$, we have
\begin{equation}\label{e101}
  \lim_{\sigma_2^2,\ldots,\sigma_M^2 \to\infty}\ I(\textbf{\textit{X}};\textbf{\textit{X}}+\textbf{\textit{Z}}|U)=I(X_1;X_1+Z_1|U)
\end{equation}
The proof is quite the same as that in \cite{extremal} (i.e., (153) to (159)) considering the following Markov chains: $X_1+Z_1\to X_1\to \tilde{\textbf{\textit{X}}}+\tilde{\textbf{\textit{Z}}}$, $X_1\to \tilde{\textbf{\textit{X}}}\to \tilde{\textbf{\textit{X}}}+\tilde{\textbf{\textit{Z}}}$ where $\tilde{\textbf{\textit{X}}}=(X_2,\ldots,X_M)^t$ and $\tilde{\textbf{\textit{Z}}}=(Z_2,\ldots,Z_M)^t$.

By eigenvalue decomposition, we have $\textit{\textbf{K}}_{Z_1}=\textit{\textbf{A}}\textit{\textbf{$\Lambda$}}_1\textbf{\textit{A}}^t$ and $\textbf{\textit{K}}_{Z_2}=\textbf{\textit{B}}\textbf{\textit{$\Lambda$}}_2\textit{\textbf{B}}^t$ where $\textit{\textbf{A}}=[\textit{\textbf{a}}_1,\ldots,\textit{\textbf{a}}_m]$, $\textit{\textbf{B}}=[\textit{\textbf{b}}_1,\ldots,\textit{\textbf{b}}_m]$ and $\Lambda_i=\mbox{diag}(\lambda_{i1},\lambda_{i2},\ldots,\lambda_{iM}) (i=1,2)$. The columns of $\textbf{\textit{A}}$ are $M$ orthogonal vectors in an $M$ dimensional space. In our derivations, these vectors do not need to be orthonormal (i.e., \textbf{\textit{A}} and \textbf{\textit{B}} are not necessarily unitary matrices), but we only restrict them to have finite norms. Since $\textbf{\textit{A}}$ and $\textbf{\textit{B}}$ are invertible (due to being full rank), we can write
\begin{equation}\label{e102}
  \lim_{\lambda_{12}^2,\ldots,\lambda_{1M}^2 \to\infty}\ I(\textbf{\textit{X}};\textbf{\textit{X}}+\textbf{\textit{Z}}_1|U)= \lim_{\lambda_{12}^2,\ldots,\lambda_{1M}^2 \to\infty}\ I(\textit{\textbf{A}}^t\textbf{\textit{X}};\textit{\textbf{A}}^t\textbf{\textit{X}}+\underbrace{\textit{\textbf{A}}^t\textbf{\textit{Z}}_1}_{\hat{\textbf{\textit{Z}}}_1}|U)
\end{equation}
where $\hat{\textbf{\textit{Z}}}_1$ is a Gaussian vector independent of $U$ with independent elements having the variances $\hat{\sigma}_j^2=\lambda_{1j}\|\textit{\textbf{a}}_j\|^4 (j=1,2,\ldots,M)$ where $\textit{\textbf{a}}_j$ is the $j^{th}$ column of $\textit{\textbf{A}}$ having a finite norm. According to (\ref{e101}),
\begin{equation}\label{e103}
  (\ref{e102})=I(\textit{\textbf{a}}_1^t\textit{\textbf{X}};\textit{\textbf{a}}_1^t\textit{\textbf{X}}+\hat{Z}_{11}|U)
\end{equation}
where $\hat{Z}_{11}$ denotes the first element of $\hat{\textbf{\textit{Z}}}_1$. Following the same steps as in \cite{extremal} (from (162) to (167)), we get
\begin{equation}\label{e104}
  h(\textit{\textbf{a}}_1^t\textit{\textbf{X}}+\hat{Z}_{11}|U)-\mu h(\textit{\textbf{b}}_1^t\textit{\textbf{X}}+\hat{Z}_{21}|U)\leq \max_{0\preceq \textit{\textbf{K}}_{X|U}\preceq \textbf{\textit{S}}}\
\left\{\frac{1}{2}\log (2\pi e(\textit{\textbf{a}}_1^t\textit{\textbf{K}}_{X|U}\textit{\textbf{a}}_1+\hat{\sigma}_{11}^2))-\frac{\mu}{2}\log (2\pi e(\textit{\textbf{b}}_1^t\textit{\textbf{K}}_{X|U}\textit{\textbf{b}}_1+\hat{\sigma}_{21}^2))\right\}
\end{equation}
where $\textit{\textbf{K}}_{X|U}=\mbox{Cov}(\textit{\textbf{X}}|U)$. \qedhere
\end{proof}
We split $T$ in (\ref{e10}) as $T=(U,\textbf{\textit{h}}_m(j),\textbf{\textit{h}}_q(j))$. From now on, we drop the time indices for simplicity. Therefore,
\begin{align}
  h(Y_m|T)-h(Y_q|T) &\leq \max_{\textbf{\textit{S}}:\textit{\textbf{S}}\succeq0, \mbox{tr}(\textbf{\textit{S}})\leq P}\ \max_{0\preceq \textbf{\textit{K}}_{X|U}\preceq \textit{\textbf{S}}}\ h(Y_m|U,\textbf{\textit{h}}_m,\textbf{\textit{h}}_q)-h(Y_q|U,\textbf{\textit{h}}_m,\textbf{\textit{h}}_q) \nonumber \\
&=  \max_{\textbf{\textit{S}}:\textit{\textbf{S}}\succeq0, \mbox{tr}(\textbf{\textit{S}})\leq P}\ \max_{0\preceq \textbf{\textit{K}}_{X|U}\preceq \textit{\textbf{S}}}\ E_{\textbf{\textit{h}}_m,\textbf{\textit{h}}_q} \left\{h(Y_m|U,\textbf{\textit{h}}_m=\textbf{\textit{h}}_m^0,\textbf{\textit{h}}_q=\textbf{\textit{h}}_q^0)-h(Y_q|U,\textbf{\textit{h}}_m=\textbf{\textit{h}}_m^0,\textbf{\textit{h}}_q=\textbf{\textit{h}}_q^0)\right\} \nonumber \\
& \leq \max_{\textbf{\textit{S}}:\textit{\textbf{S}}\succeq0, \mbox{tr}(\textbf{\textit{S}})\leq P}\ E_{\textbf{\textit{h}}_m,\textbf{\textit{h}}_q} \left\{\max_{0\preceq \textbf{\textit{K}}_{X|U}\preceq \textit{\textbf{S}}}\ h(Y_m|U,\textbf{\textit{h}}_m=\textbf{\textit{h}}_m^0,\textbf{\textit{h}}_q=\textbf{\textit{h}}_q^0)-h(Y_q|U,\textbf{\textit{h}}_m=\textbf{\textit{h}}_m^0,\textbf{\textit{h}}_q=\textbf{\textit{h}}_q^0)\right\} \label{x2}\\
&= \max_{\textbf{\textit{S}}:\textit{\textbf{S}}\succeq0, \mbox{tr}(\textbf{\textit{S}})\leq P}\ E_{\textbf{\textit{h}}_m,\textbf{\textit{h}}_q} \left\{\max_{0\preceq \textbf{\textit{K}}_{X|U}\preceq \textit{\textbf{S}}}\ h({\textbf{\textit{h}}_m^0}^H\textit{\textbf{X}}(\textbf{\textit{h}}_m^0) + w_m|U)-h({\textbf{\textit{h}}_q^0}^H\textit{\textbf{X}}(\textbf{\textit{h}}_m^0) + w_q|U)\right\} \label{x3} \\
&\leq \max_{\textbf{\textit{S}}:\textit{\textbf{S}}\succeq0, \mbox{tr}(\textbf{\textit{S}})\leq P}\ E_{\textbf{\textit{h}}_m,\textbf{\textit{h}}_q} \left\{\max_{0\preceq \textbf{\textit{K}}_{X|U}\preceq \textit{\textbf{S}}}\ \left\{\frac{1}{2}\log (2\pi e({\textit{\textbf{h}}_m^0}^H\textit{\textbf{K}}_{X|U}\textit{\textbf{h}}_m^0+1))-\frac{\mu}{2}\log (2\pi e({\textit{\textbf{h}}_q^0}^H\textit{\textbf{K}}_{X|U}\textit{\textbf{h}}_q^0+1))\right\}\right\} \label{x4}
\end{align}
where $\textbf{\textit{h}}_m^0$ and $\textbf{\textit{h}}_q^0$ are two realizations of the random vectors $\textbf{\textit{h}}_m$ and $\textbf{\textit{h}}_q$. In (\ref{x2}) taking the maximization into the expectation makes the value greater. In (\ref{x3}), we have used the realizations and also consider that since the CSIT of $\textit{\textbf{h}}_m$ is perfect, the transmitted signal can be a function of it which is denoted by $\textit{\textbf{X}}(\textbf{\textit{h}}_m^0)$. (\ref{x4}) results from lemma 4.

Let $\lambda_1\geq \lambda_2\geq \ldots\geq \lambda_M$ be the eigenvalues of the covariance matrix $\textit{\textbf{K}}_{X|U}$. We can write
\begin{equation}\label{yek}
  \log ({\textit{\textbf{h}}_m^0}^H\textit{\textbf{K}}_{X|U}\textit{\textbf{h}}_m^0+1)\leq \log (\lambda_1\|\textit{\textbf{h}}_m^0\|^2+1)
\end{equation}
and it is achieved when the eigenvector corresponding to the largest eigenvalue of $\textit{\textbf{K}}_{X|U}$ is aligned with $\textit{\textbf{h}}_m^0$. This is in fact possible, since the transmitter has a perfect knowledge of $\textit{\textbf{h}}_m^0$ and can construct the covariance matrix (via precoding) to meet this condition. However, for user $q$, we have
\begin{equation}\label{do}
  \log ({\textit{\textbf{h}}_q^0}^H\textit{\textbf{K}}_{X|U}\textit{\textbf{h}}_q^0+1)= \log (\sum_{i=1}^M \lambda_i|\alpha_i|^2+1)
\end{equation}
where $\alpha_i$ is the projection of $i^{th}$ eigenvector of $\textit{\textbf{K}}_{X|U}$ onto $\textit{\textbf{h}}_q^0$ (i.e., $\alpha_i={\textit{\textbf{h}}_q^0}^H\textit{\textbf{v}}_i$, where $\textit{\textbf{v}}_i$ is the eigenvector corresponding to $\lambda_i$.) Since the transmitter has no CSIT of user $q$, the transmitted signal will be independent of $\textit{\textbf{h}}_q$. This results in having $\alpha_1\neq 0$ almost surely (or equivalently, $\mbox{Pr}\{\alpha_1=0\}=0$.) In other words, if $E$ denotes the event of having $\textit{\textbf{h}}_q$ orthogonal to the eigenvector corresponding to the largest eigenvalue of $\textit{\textbf{K}}_{X|U}$, then the dimension of $E$ is lower than the dimension of the sample space resulting in $Pr\{E\}=0$. By replacing $\mu$ with $1$ in (\ref{x4}) and using (\ref{yek}) and (\ref{do}), we get
\begin{align}
  \log ({\textit{\textbf{h}}_m^0}^H\textit{\textbf{K}}_{X|U}\textit{\textbf{h}}_m^0+1)-\log ({\textit{\textbf{h}}_q^0}^H\textit{\textbf{K}}_{X|U}\textit{\textbf{h}}_m^0+1)&\leq \log (\frac{\lambda_1\|\textit{\textbf{h}}_m^0\|^2+1}{\lambda_1|\alpha_1|^2+1})\\
&\leq \frac{\|\textit{\textbf{h}}_m^0\|^2\lambda_1}{1+\|\textit{\textbf{h}}_m^0\|^2\lambda_1}\log(\frac{\|\textit{\textbf{h}}_m^0\|^2}{|\alpha_1|^2})\label{76}\\
&\leq \underbrace{\log(\frac{\|\textit{\textbf{h}}_m^0\|^2}{|\alpha_1|^2})}_{O(\log P)} \label{77}
\end{align}
where (\ref{76}) results from the application of log sum inequality \cite[p. 30]{Cover} and (\ref{77}) is due to the fact that $\|\textit{\textbf{h}}_m^0\|$ and $|\alpha_1|$ do not scale with $P$. Since the DoF analysis is in infinite SNR regime, the exact value of $|\alpha_1|$ is not important as long as it is non-zero. Therefore, for each realization of the channel, inside the expectation of (\ref{x4}) is of order $O(\log P)$, and so will be the expectation. Hence, when CSIT of $\textit{\textbf{h}}_m$ is perfect and CSIT of $\textit{\textbf{h}}_q$ is not known
\begin{equation}
  \lim_{P\to \infty} \frac{h(Y_m|T)-h(Y_q|T)}{\log P}=0.
\end{equation}
This completes the proof.
\section{An alternative proof of $\sum_{i=1}^K\frac{d_i}{i}\leq 1 + \sum_{i=2}^K\frac{\sum_{r=1}^{i-1}\lambda_P^r}{i(i-1)}$}\label{s4}
The proof is based on the approach used in \cite{Gesbert}, therefore the following definitions are necessary. The channel vector of user $k$ at time $n$ can be written as
\begin{equation}
\textbf{\textit{h}}_k(n)=\widehat{\textbf{\textit{h}}}_k(n) + \widetilde{\textbf{\textit{h}}}_k(n)
\end{equation}
where $\widehat{\textbf{\textit{h}}}_k(n)$ and $\widetilde{\textbf{\textit{h}}}_k(n)$ are the estimate of the channel and estimation error with distributions $CN({\textbf{0}},(1-\sigma_{k}^{2}(n))\textbf{I})$ and $CN({\textbf{0}},\sigma_{k}^{2}(n)\textbf{I})$, respectively.  The variance of error is

\[ \sigma_{k}^{2}(n) = E\left[\|\widetilde{\textbf{\textit{h}}}_k(n)\|^{2}\right]. \]
As observed from the above, although the channel is assumed stationary, the estimate is a non-stationary process meaning that the quality of estimation varies over time. The quality of CSIT for user $k$ at time instant $n$ is
\begin{equation}
\alpha_k(n)=-\lim_{P \to \infty}\frac{\log{\left(\sigma_{k}^{2}(n)\right)}}{\log{P}}.
\end{equation}
From the results of \cite{Jindal}, if the rate of feedback scales linearly with $\log P$ (or equivalently, the variance of estimation error decrease as $O(P^{-1})$ or faster), perfect CSIT multiplexing gain can be obtained. Therefore, the effective range of $\alpha_k(n)$ will be $[0,1]$ where in terms of DoF, $\alpha_k(n)=1$ could be interpreted as perfect CSIT of user $k$ at time instant $n$. We also define $\widehat{H}(n)=[\widehat{\textbf{\textit{h}}}_1(n),\ldots,\widehat{\textbf{\textit{h}}}_K(n)]^H$  and ${\widehat{H}}^n = \{\widehat{H}(1),\ldots,\widehat{H}(n)\}$. Again, for simplicity, we show the inequalities for a fixed permutation of the users while the results could be easily extended to any arbitrary permutations.
As in part \emph{A} of the first proof, the same channel improvement is done here. The only difference is that we assume the users not only have perfect global CSIR, but also they know the channel estimates at the transmitter. Applying this difference to formulae (\ref{Fano}) to (\ref{e30}), we rewrite (\ref{e30}) as
\begin{multline}\label{ee30}
\sum_{i=1}^K\frac{nR_i}{i}  \leq\overbrace{h(Y_1^n|H^n,\widehat{H}^n)}^{\leq n\log P} + \\
\sum_{i=2}^K\sum_{j=1}^n\left[\frac{h(Y_1(j),\ldots,Y_i(j)|T_{i,j},H(j))}{i}-
 \frac{h(Y_1(j),\ldots,Y_{i-1}(j)|T_{i,j},H(j))}{i-1}\right] + nO(\log P).
\end{multline}
where $T_{i,j}=(W_1,\ldots,W_{i-1},Y_1^{j-1},\ldots,Y_i^{j-1},H^{j-1},\widehat{H}^j)$.
In what follows, we find an upper bound for the term in the brackets of (\ref{ee30}). Following the same approach as in \cite{Gesbert}, we can write
\begin{align}
   &\max_{P_{T_{i,j}}P_{\textbf{\textit{x}}(j)|T_{i,j}}}\left[\frac{h(Y_1(j),\ldots,Y_i(j)|T_{i,j},H(j))}{i}- \frac{h(Y_1(j),\ldots,Y_{i-1}(j)|T_{i,j},H(j))}{i-1}\right] \\
   &\leq \max_{P_{T_{i,j}}}E_{{T_{i,j}}}\left[\max_{P_{\textbf{\textit{x}}(j)|T_{i,j}}}\left(\frac{h(Y_1(j),\ldots,Y_i(j)|T_{i,j}=T,H(j))}{i}-
   \frac{h(Y_1(j),\ldots,Y_{i-1}(j)|T_{i,j}=T,H(j))}{i-1}\right)\right]
\end{align}
\begin{align}
   &= \max_{P_{T_{i,j}}}E_{{T_{i,j}}}\left[\max_{P_{\textbf{\textit{x}}(j)|T_{i,j}}}E_{H(j)|{T_{i,j}}}\left(\frac{h(Y_1(j),\ldots,Y_i(j)|T_{i,j}=T,H(j)=H)}{i}-
   \frac{h(Y_1(j),\ldots,Y_{i-1}(j)|T_{i,j}=T,H(j)=H)}{i-1}\right)\right]\\
   &= \max_{P_{T_{i,j}}}E_{{T_{i,j}}}\left[\max_{P_{\textbf{\textit{x}}(j)|T_{i,j}}}E_{H(j)|\widehat{H}(j)}\left(\frac{h(H_1^i(j)\textbf{\textit{x}}(j)+\textbf{\textit{n}}_j|T_{i,j}=T)}{i}-
   \frac{h(H_1^{i-1}(j)\textbf{\textit{x}}(j)+\textbf{\textit{w}}_j|T_{i,j}=T)}{i-1}\right)\right]\\
   &= \max_{P_{T_{i,j}}}E_{{T_{i,j}}}\left[\max_{\textbf{\textit{C}}:\textbf{\textit{C}}\succeq 0,tr(\textbf{\textit{C}})\leq P}\max_{\substack{P_{\textbf{\textit{x}}(j)|T_{i,j}}\\Cov(\textbf{\textit{x}}(j)|T_{i,j})\preceq \textit{\textbf{C}}}}E_{H(j)|\widehat{H}(j)}\left(\frac{h(H_1^i(j)\textbf{\textit{x}}(j)+\textbf{\textit{n}}_j|T_{i,j}=T)}{i}-
   \frac{h(H_1^{i-1}(j)\textbf{\textit{x}}(j)+\textbf{\textit{m}}_j|T_{i,j}=T)}{i-1}\right)\right]\\
   &= \max_{P_{T_{i,j}}}E_{{T_{i,j}}}\left[\max_{\textbf{\textit{C}}:\textbf{\textit{C}}\succeq0,tr(\textbf{\textit{C}})\leq P}E_{H(j)|\widehat{H}(j)}\left(\frac{\log\det{(I_i+H_1^i(j)\textbf{K}_*H_1^i(j)^H)}}{i}-
   \frac{\log\det{(I_{i-1}+H_1^{i-1}(j)\textbf{K}_*H_1^{i-1}(j)^H)}}{i-1}\right)\right] \label{ee1}\\
   &\leq E_{\widehat{H}(j)}\left[\max_{\textbf{\textit{K}}:\textbf{\textit{K}}\succeq0,tr(\textbf{\textit{K}})\leq P} E_{H(j)|\widehat{H}(j)} \left(\frac{\log\det{(I_i+H_1^i(j)\textbf{K}H_1^i(j)^H)}}{i}
  -\frac{\log\det{(I_{i-1}+H_1^{i-1}(j)\textbf{K}H_1^{i-1}(j)^H)}}{i-1}\right)\right] \\
   &\leq  -\frac{\log\det{(\Sigma^2)}}{i(i-1)}+O(\log P) \label{eee}
\end{align}
where we have the Markov chain $\textbf{x}(j)\leftrightarrow T_{i,j}\leftrightarrow\hat{H}(j)\leftrightarrow H(j)$, $H_1^i(j)=\left[\textbf{\textit{h}}_1(j),\ldots,\textit{\textbf{h}}_i(j)\right]^H$,\ $H_1^{i-1}(j)=\left[\textbf{\textit{h}}_1(j),\ldots,\textit{\textbf{h}}_{i-1}(j)\right]^H$, $\textbf{\textit{n}}_j=\left[w_1(j),\ldots,w_i(j)\right]^T$ and $\textbf{\textit{m}}_j=\left[w_1(j),\ldots,w_{i-1}(j)\right]^T$. (\ref{ee1}) is the application of extremal inequality \cite{extremal}, \cite{weingarten} where the Gaussian distribution maximizes a specific difference between two differential entropies. The last inequality (\ref{eee}) comes from (101) in \cite{Elia}, in which
\begin{equation*}
  \Sigma^2=diag\left(\sigma_1^2(j),\ldots,\sigma_{i-1}^2(j)\right).
\end{equation*}
Therefore, we can write
\begin{eqnarray}
 \sum_{i=1}^K\frac{nR_i}{i}&\leq& n\log P +
 \sum_{i=2}^K\sum_{j=1}^n\left[-\frac{\log\det{(\Sigma^2)}}{i(i-1)}+O(\log P)\right ] +nO(\log P) \nonumber \\
 &=& n\log P + \sum_{i=2}^K\frac{\sum_{j=1}^n[\alpha_1(j)+\cdots+\alpha_{i-1}(j)]}{i(i-1)}\log P
 + nKO(\log P).
\end{eqnarray}
Since the channel is degraded and $D$ is replaced with $N$, the CSIT is either $P$ or $N$. Therefore,  the $\alpha$'s are either $1$ with probability $\lambda_P^i$ or $0$ otherwise. Therefore, for $n$ large enough, we have
\begin{equation*}
  \lim_{n\to \infty}\sum_{j=1}^n[\alpha_1(j)+\cdots+\alpha_{i-1}(j)]=n\sum_{r=1}^{i-1}\lambda_P^r
\end{equation*}
which results in
\begin{equation}\label{e2}
  \sum_{i=1}^K\frac{nR_i}{i}\leq n\log P + \sum_{i=2}^K\sum_{r=1}^{i-1}\frac{n\lambda_P^r}{i(i-1)}\log P
  + nKO(\log P)
\end{equation}
at large $n$. Dividing both sides by $n\log P$ and taking the limit of (\ref{e2}) as $n,P\to \infty$, we get
\begin{equation}
 \sum_{i=1}^K\frac{d_i}{i}\leq 1 + \sum_{i=2}^K\frac{\sum_{r=1}^{i-1}\lambda_P^r}{i(i-1)}.
\end{equation}
It is obvious that the same approach can be applied to any other permutations of $(1,2,\ldots,K)$.
\section{Proof of (\ref{e33})}\label{s5}
According to \textit{Csisz\'{a}r sum identity} \cite{network_info}, for the two arbitrary random vectors $X^n$ and $Y^n$
\begin{equation}\label{e6e}
  \sum_{i=1}^n I(X_{i+1}^n;Y_i|Y^{i-1})=\sum_{i=1}^n I(Y^{i-1};X_i|X_{i+1}^n).
\end{equation}
where $X_{n+1},Y_0=\emptyset$. By writing the mutual information of (\ref{e6e}) in terms of the differential entropies, we have
\begin{equation}\label{e7e}
\sum_{i=1}^n [h(Y_i|Y^{i-1})-h(Y_i|Y^{i-1},X_{i+1}^n)]=\sum_{i=1}^n [h(X_i|X_{i+1}^n)-h(X_i|X_{i+1}^n,Y^{i-1})].
\end{equation}
and finally, by using the chain rule of entropies, we get
\begin{equation}
  h(X^n)-h(Y^n) = \sum_{i=1}^n [h(X_i|X_{i+1}^n,Y^{i-1})-h(Y_i|X_{i+1}^n,Y^{i-1})]
\end{equation}
\bibliography{REFERENCE}
\bibliographystyle{IEEEtran}
\end{document}